\documentclass[11pt]{article}
\usepackage{fullpage}
\usepackage{amsmath}
\usepackage{amsfonts}

\newcounter{algoctr}

\newenvironment{algo}{\begin{list}{\arabic{algoctr}.}{\usecounter{algoctr}
        \setlength{\parsep}{0pt} \setlength{\itemsep}{0pt}}}{\end{list}}

\newcommand{\algitem}[1]{\item \hspace*{#1 em}}

\newif\ifnotesw\noteswtrue
   {\ifnotesw\marginpar[\hfill\(\top\)]{\(\top\)}\fi}%
   {\ifnotesw\marginpar[\hfill\(\bot\)]{\(\bot\)}\fi}

\newcommand{\mnote}[1]%
    {\ifnotesw\marginpar%
        [{\scriptsize\begin{minipage}[t]{\marginparwidth}
        \raggedleft#1%
                        \end{minipage}}]%
        {\scriptsize\begin{minipage}[t]{\marginparwidth}
        \raggedright#1%
                        \end{minipage}}%
    \fi}

\newcommand{\ignore}[1]{}

\newcommand{\szon}{\mbox{$\{0,1\}^n$}}

\newcommand{\smp}{\mbox{$\{-1,+1\}$}}

\newcommand{\sign}{\mbox{\rm sign}}

\newsavebox{\given}
\savebox{\given}[1em]{\rule[-1.5ex]{.2mm}{4ex}}

\newcommand{\Fo}{\mbox{\tt F1}}
\newcommand{\BComb}{\mbox{\tt B$_{\mbox{Comb}}$}}

\newcommand{\bnum}{\begin{equation}}
\newcommand{\enum}{\end{equation}}

\newtheorem{theorem}{Theorem}

\newtheorem{lemma}[theorem]{Lemma}

\def\Exp {\mbox{\bf E}}

\newcommand{\blackslug}{\rule{7pt}{7pt}}

\newcommand{\iverson}[1]{\lbrack\!\lbrack #1 \rbrack\!\rbrack}

\newcommand{\real}{\ifmmode {\rm R} \else ${\rm R}$ \fi}
\newcommand{\nat}{\ifmmode {\rm N} \else ${\rm N}$  \fi}
\newcommand{\tot}{\ifmmode {\cal T} \else ${\cal T}$ \fi}
\newcommand{\sigstar}{\ifmmode \Sigma^{\ast} \else $\Sigma^{\ast}$ \fi}

\newcommand{\inn}{\ifmmode \in \else $\in$ \fi}
\renewcommand{\phi}{\ifmmode \varphi \else $\varphi$ \fi}
\renewcommand{\le}{\ifmmode \leq \else $\leq$ \fi}
\renewcommand{\ge}{\ifmmode \geq \else $\geq$ \fi}
\renewcommand{\ne}{\ifmmode \neq \else $\neq$ \fi}
\newcommand{\lt}{\ifmmode < \else $<$ \fi}
\newcommand{\gt}{\ifmmode > \else $>$ \fi}
\newcommand{\eq}{\ifmmode = \else $=$ \fi}
\newcommand{\half}{\ifmmode \frac{1}{2} \else $\frac{1}{2}$ \fi}
\newcommand{\oneovern}{\ifmmode \frac{1}{n} \else $\frac{1}{n}$ \fi}

\newcommand{\ra}{\ifmmode \rightarrow \else $\rightarrow$ \fi}
\newcommand{\qed}{\hfill{\setlength{\fboxsep}{0pt}
\framebox[7pt]{\rule{0pt}{7pt}}}}

\renewcommand{\notin}{\ifmmode \not\in \else $\not\in$ \fi}
\newlength{\thislabel}
\newcommand{\labsize}[1]{\settowidth{\thislabel}{#1}}

\newcommand{\prf}{\par\noindent{\sl Proof } \hspace{.01 in}}
\newcommand{\zo}{\{0,1\}}

\newcommand{\lip}{\langle}
\newcommand{\rip}{\rangle}

\def\Real{\mathbb{R}}

\def\dotdef{\stackrel{\cdot}{=}}

\title{Quantum DNF Learnability Revisited\\
        {\small (preliminary version)}}
\author{{\em Jeffrey C. Jackson}\thanks{This material is based upon work 
        supported by the National Science Foundation under Grant
        No.\ CCR-9877079.} \\ Duquesne University
        \and 
        {\em Christino Tamon}\thanks{This research was supported by the 
	National Science Foundation Grant No.\ DMR-0121146.} \\ Clarkson University
        \and 
	{\em Tomoyuki Yamakami} \\ University of Ottawa}
\date{\today}   

\begin{document}
\bibliographystyle{plain}
\maketitle

\begin{abstract}
  We describe a quantum PAC learning algorithm for DNF formulae under
  the uniform distribution with a query complexity of
  $\tilde{O}(s^{3}/\epsilon + s^{2}/\epsilon^{2})$, where $s$ is the
  size of DNF formula and $\epsilon$ is the PAC error accuracy.  If
  $s$ and $1/\epsilon$ are comparable, this
  gives a modest improvement over a previously known classical query
  complexity of $\tilde{O}(ns^{2}/\epsilon^{2})$.  We also show a
  lower bound of $\Omega(s\log n/n)$ on the query complexity of any
  quantum PAC algorithm for learning a DNF of size $s$ with $n$ inputs
  under the uniform distribution.
\end{abstract}

\section{Introduction}

In this abstract we describe a quantum learning algorithm for DNF formulae under the uniform distribution
using quantum membership queries. Although Bshouty and Jackson \cite{bj99} have shown that it is possible
to adapt Jackson's Harmonic Sieve algorithm \cite{j97} to the quantum setting, our goal is different.
We will focus on reducing the number of quantum membership queries used by the DNF learning algorithm 
whereas their motivation was in showing that quantum examples are sufficient for learning DNF. 

The Harmonic Sieve {\tt HS} algorithm combines two crucial independent algorithms. The first algorithm 
is an inner algorithm for finding parity functions that weakly approximate the target DNF function. 
The second algorithm used in the Harmonic Sieve is an outer algorithm that is a boosting algorithm. 
A weak learning algorithm is an algorithm that produces hypotheses whose accuracy are slightly
better than random guessing. Boosting is a method for improving the accuracy of hypotheses given
by a weak learning algorithm. 

For the inner algorithm, a Fourier-based algorithm given in \cite{km93} (called the KM algorithm) 
is used in {\tt HS} for finding the weak parity approximators. 
The KM algorithm is based on a similar method given by Goldreich and Levin \cite{gl89} in their 
seminal work on hardcore bits in cryptography. 
Subsequently, Levin \cite{l93} and Goldreich \cite{g}, independently, gave highly improved methods for 
solving this so-called {\em Goldreich-Levin} problem. Their ideas were adapted by Bshouty et al. \cite{bjt99} 
to obtain a weak DNF learning algorithm with query and time complexity of $\tilde{O}(n/\gamma^2)$,
where $\gamma$ is the weak advantage of the parity approximator. By a result of Jackson \cite{j97},
$\gamma = O(1/s)$ for DNF formula of size $s$.

For the outer algorithm, the original {\tt HS} used a boosting method of Freund \cite{f95} called $\Fo$ 
that has various nice features. Recently, Klivans and Servedio \cite{ks99} observed that a construction
of Impagliazzo \cite{imp98} gave a {\em smoother} boosting algorithm called {\tt IHA}.
It was shown that {\tt IHA} is a $O(1/\epsilon)$-smooth $O(\gamma^{-2}\epsilon^{-2})$-stage boosting 
algorithm, where $\gamma$ is the weak advantage of the weak learning algorithm and $\epsilon$ is the 
target accuracy. In contrast, {\tt F1} is a $O(1/\epsilon^3)$-smooth $O(\gamma^{-2}\log(1/\epsilon))$-stage
boosting algorithm.

The fastest known algorithm for learning DNF is obtained by combining the two improved independent 
components that results in a total running time of $\tilde{O}(ns^{4}/\epsilon^{2})$ and a query
complexity of $\tilde{O}(ns^{2}/\epsilon^{2})$ \cite{bjt99,ks99}.

We describe an efficient {\em quantum} DNF learning algorithm by combining a quantum Goldreich-Levin
algorithm {\tt QGL} of Adcock and Cleve \cite{ac01} with a well-known highly efficient boosting algorithm 
of Freund called $\BComb$ \cite{f95}. The quantum algorithm of Adcock and Cleve used only $O(1/\gamma)$ 
queries (beating a classical lower bound of $\Omega(n/\gamma^2)$ proved also in \cite{ac01}).
Freund's $\BComb$ algorithm is a $\tilde{O}(1/\epsilon)$-smooth $O(\gamma^{-2}\log(1/\epsilon))$-stage
boosting algorithm. After adapting both algorithms for quantum PAC learning, we obtain a quantum Harmonic 
Sieve algorithm {\tt QHS} with a sample complexity of $\tilde{O}(s^3/\epsilon + s^{2}/\epsilon^{2})$.
In contrast to the best known classical upper bound of $\tilde{O}(ns^2/\epsilon^2)$, this gives a
modest improvement if $s$ and $1/\epsilon$ are comparable.

As shown in \cite{ac01}, the quantum Goldreich-Levin algorithm has applications to quantum
cryptography. In this work, we show one of its applications in computational learning theory.

For the sake of exposition, in this abstract we will describe our quantum DNF PAC learning algorithm
using a conceptually simpler boosting algorithm {\tt SmoothBoost} given by Servedio \cite{s01}. We
describe a {\em boost-by-filtering} version of Servedio's {\tt SmoothBoost} that is a
$O(1/\epsilon)$-smooth $O(\gamma^{-2}\epsilon^{-1})$-stage boosting algorithm. 
So, we incur an extra $1/\epsilon$ factor in the sample complexity. 
We defer the details of using $\BComb$ in {\tt QHS} to the final version of this paper.

Finally, we prove a query lower bound of $\Omega(s\log n/n)$ on any quantum PAC learning algorithm
for DNF under the uniform distribution with (quantum) membership queries. 

\ignore{
We also discuss how this new quantum Harmonic Sieve compares with Bshouty-Jackson's query-free quantum
algorithm \cite{bj99}.
}

\section{Preliminaries}

We are interested in algorithms for learning approximations to an
unknown function that is a member of a particular class of functions.
The specific function class of interest in this paper is that of DNF
expressions, that is, Boolean functions that can be expressed as a
disjunction of terms, where each term is a conjunction of Boolean
variables (possibly negated).  Given a {\em target} DNF expression $f
: \szon \rightarrow \smp$ having $s$ terms along with an {\em
  accuracy} parameter $0 < \epsilon < 1/2$
  and a {\em confidence} parameter $\delta > 0$,
the goal is to with probability at least $1-\delta$ produce
a {\em hypothesis} $h$ such $\Pr_{x \sim U_n}[f(x) \neq h(x)] <
\epsilon$, where $U_n$ represents the uniform distribution over
$\szon$.  We will sometimes refer to such an $h$ as an
$\epsilon$-approximator to $f$, or equivalently say that $h$ has $\half
- \epsilon$ {\em advantage} (this represents the advantage over the
agreement between $f$ and a random function, which is $1/2$).
A learning algorithm that can guarantee only $\gamma > 0$ advantage
in the hypothesis produced but can do so with arbitrarily small probability
of failure $\delta$ 
is called a {\em weak learning algorithm},
and the hypothesis produced is a {\em weak approximator}.

The information our learning algorithm is given about the target
function varies.  One form is a {\em sample}, that is, a set $S$ of
input/output pairs for the function.  We often use $x$ to denote an
input and $f(x)$ the associated output, and $x \in S$
to denote that $x$ is one of the inputs of the pairs in $S$.
Another type of information we sometimes use is a {\em membership oracle}
for $f$, $MEM_f$.  Such an oracle is given an input $x$ and returns
the function's output $f(x)$.
\ignore{
Our primary source of information about $f$ will be a
{\em quantum membership oracle} $QMQ_f$.
}

\section{A smoother Boost-by-Filtering algorithm}

\begin{figure} 

\begin{tabbing}
{\bf Output:} \= www \= www \= \kill
{\bf Input:} \> Parameters  $0 < \epsilon < 1/2$, $0 \le \gamma < 1/2$ \\
             \> Sample $S$ of target $f$ \\
             \> Weak learning algorithm {\tt WL} \\
\\
{\bf Output:} \> Hypothesis $h$
\end{tabbing}
\begin{algo}
\algitem{0} $U_S$ $\equiv$ the uniform distribution over $S$
\algitem{0} $M_{1}(x) \equiv 1$, $\forall x \in S$
\algitem{0} $N_{0}(x) \equiv 0$, $\forall x \in S$
\algitem{0} $\theta \leftarrow \gamma / (2 + \gamma)$
\algitem{0} $t \leftarrow 1$
\algitem{0} {\bf while } $\Exp_{x \sim U_S}[M_{t}(x)] > \epsilon$ {\bf do} \label{lin:sbwhile}
\algitem{1} $D_{t}(x) \equiv M_{t}(x)/(m\Exp_{x \sim U_S}[M_{t}(x)])$,
$\forall x \in S$ \label{lin:sbdt}
\algitem{1} $h_{t} \leftarrow \mathtt{WL}(S,D_{t},
\delta = \Omega(\epsilon^{-1} \gamma^{-2}))$
\algitem{1} $N_{t}(x) \equiv N_{t-1}(x) + f(x)h_{t}(x) - \theta$, $\forall x \in S$
\algitem{1} $M_{t+1}(x) \equiv \iverson{N_{t}(x) < 0} + (1-\gamma)^{N_{t}(x)/2}\iverson{N_{t}(x) \ge 0}$,
        $\forall x \in S$
\algitem{1} $t  \leftarrow t + 1$
\algitem{0} {\bf end while}
\algitem{0} $T \leftarrow t - 1$
\algitem{0} $H \equiv \frac{1}{T}\sum_{i=1}^{T} h_{i}$
\algitem{0} {\bf return} $h \equiv \sign(H)$.
\end{algo}
\hrule

\caption[Servedio's SmoothBoost algorithm.]
{The {\tt SmoothBoost} algorithm of Servedio \cite{s01}.}
\label{figure:smoothboost}
\end{figure}

A modification of Servedio's {\tt SmoothBoost} boosting algorithm
\cite{s01} is described in this section.  A special case (discrete
weak hypotheses and fixed margin) version of {\tt SmoothBoost} sufficient
for our purposes is shown in 
Figure~\ref{figure:smoothboost}.
{\tt SmoothBoost} is a boosting-by-sampling method
that can be applied to a weak learning algorithm in order to produce a
hypothesis that closely approximates the sample.  Specifically, {\tt
  SmoothBoost} receives as input a sample $S$ of size $m$ as well as
accuracy parameter $\epsilon$.  It is also given a weak learning
algorithm {\tt WL}.  The boosting algorithm defines a series of
distributions $D_t$ over $S$ and successively calls the weak learning
algorithm, providing it with the sample $S$ and with one of the
distributions $D_t$.  In the end, the algorithm combines the
weak hypotheses returned by the calls to the weak learner into
a single hypothesis $h$.

Servedio proves three key properties of {\tt SmoothBoost}:

\begin{lemma}[Servedio] \label{lem:smoothboost}
  Let $f$ be a target function, and let $S$, $\epsilon$, $\gamma$, $h$,
  and $D_{t}$ be as defined in Figure~\ref{figure:smoothboost}.  Then
\begin{enumerate}
\item If every weak hypothesis $h_t$ returned by {\tt WL}
has advantage at least $\gamma$ with respect to $D_t$, 
 then {\tt SmoothBoost} will terminate after
        $T = O(\epsilon^{-1}\gamma^{-2})$ stages.
\item If {\tt SmoothBoost} terminates, then
$\Pr_{x \sim U_S}[f(x) \neq h(x)] < \epsilon$, where $U_S$
represents the uniform distribution over $S$ (Servedio actually proves
a stronger margin result that implies this).
\item $L_{\infty}(m D_{t}) \le 1/\epsilon$ for all $t$, where $m = |S|$
(this is the {\em smoothness property} of {\tt SmoothBoost}).
\end{enumerate}
\end{lemma}

\ignore{
\prf
Note that $D_{t}(x) = M_{t}(x)/(2^{n}\Exp[M_{t}]) \le 1/(2^{n}\epsilon)$ since
$\Exp[M_{t}] \ge \epsilon$ is an invariant during the boosting stages.
This proves $L_{\infty}(2^{n}D_{t}) \le 1/\epsilon$.

Note that $N_{T}(x) = \sum_{t=0}^{T} (f(x)h_{t}(x)-\theta) = T(f(x)H(x)-\theta)$.
Thus, if $f(x)H(x) \le \theta$, then $N_{T}(x) \le 0$ which implies $M_{T+1}(x) = 1$.
Therefore, $\Pr[f(x)H(x) \le \theta] \le \Exp[M_{T+1}] < \epsilon$ by the stopping condition.
This proves the second claim.

To prove the third claim, we follow \cite{s01} by breaking it into two subclaims. First, we prove that
\begin{equation} \label{eqn:claim1}
\Exp[\sum_{t=1}^{T} M_{t}(x)f(x)h_{t}(x)] \ge 2\gamma\sum_{t=1}^{T} \Exp[M_{t}(x)].
\end{equation} 
Then, we prove that if $\theta = \gamma/(2+\gamma)$ then
\begin{equation} \label{eqn:claim2}
\Exp[\sum_{t=1}^{T} M_{t}(x)f(x)h_{t}(x)] 
        < \frac{2}{\gamma\sqrt{1-\gamma}} + \gamma\sum_{t=1}^{T}\Exp[M_{t}(x)].
\end{equation}
Note that by combining (\ref{eqn:claim1}) and (\ref{eqn:claim2}), we obtain
\[
        \sum_{t=1}^{T} \Exp[M_{t}(x)] < \frac{2}{\gamma^{2}\sqrt{1-\gamma}}
\]
which yields $T < 2\epsilon^{-1}\gamma^{-2}$, since $\Exp[M_{t}(x)] > \epsilon$ and $\gamma < 1/2$.

Now we prove Equations (\ref{eqn:claim1}) and (\ref{eqn:claim2}). 
To see (\ref{eqn:claim1}), note that $f(x)h_{t}(x) = 1 - |f(x) - h_{t}(x)|$.
Thus $\Exp_{D_{t}}[f(x)h_{t}(x)] = 1 - \Exp_{D_{t}}[|f(x)-h_{t}(x)|] \ge 2\gamma$, by the weak
learning assumption. This implies that 
\[
        \Exp[M_{t}(x)f(x)h_{t}(x)] \ge 2\gamma\Exp[M_{t}(x)]
\]
After summing over $t$, we get
\[
        \sum_{t=1}^{T}\Exp[M_{t}(x)f(x)h_{t}(x)] \ge 2\gamma\sum_{t=1}^{T}\Exp[M_{t}(x)].
\]
as required. The proof of (\ref{eqn:claim2}) follows \cite{s01} and is omitted from this abstract.
\qed

}

Here we adapt this algorithm to obtain a boosting-by-filtering
algorithm that will be used by the Harmonic Sieve.  First, notice that
Lemma~\ref{lem:smoothboost} holds for the special case $S = \szon$.
However, there are potential problems with running the SmoothBoost
algorithm directly on such a large $S$.  First, it is not
computationally feasible to exactly compute $\Exp_{x \sim U_n}[M_t(x)]$,
where $U_n$ represents the uniform distribution over $\szon$.
So instead we must estimate this quantity by sampling.  This has a
small impact on both the form of the loop condition for the algorithm
(line~\ref{lin:sbwhile}), but also on the ``distributions'' $D_t$
passed to the weak learner (line~\ref{lin:sbdt}).  In fact, the
$D_t$ that will be passed to the weak learner will generally
not be a true distribution at all, but instead a constant multiplied
by a distribution due to the constant error in our estimate of
$\Exp_{x \sim U_n}[M_t(x)]$.

We will deal with the weak learner later, so for now let us assume
that the weak learner produces the same hypothesis $h_t$ given an
approximation to $D_t$ as it would given the actual distribution.
Then notice that the computations for $N_t$ and $M_{t+1}$ are
unchanged, so the only impact on the boosting algorithm has to do with
the loop condition at line~\ref{lin:sbwhile}.  This is easily
addressed: let $E_t$ represent an estimate of $\Exp_{x \sim
  U_n}[M_t(x)]$ to within additive error $\epsilon/3$ and change
the loop condition to $E_t > 2\epsilon/3$.  Then if the loop
terminates it must be that $\Exp_{x \sim U_n}[M_t(x)]
\leq \epsilon$, as before.  It is easily verified that given
this condition, Servedio's proof implies that $h$ is an
$\epsilon$-approximator to $f$ with respect to the uniform
distribution.  Furthermore, since $\Exp_{x \sim U_n}[M_t(x)]
\geq \epsilon/3$ if the algorithm terminates, the other statements
of Lemma~\ref{lem:smoothboost} change only by constant factors.
In particular, the smoothness condition of the lemma now becomes
$L_{\infty}(2^n D_t) \leq 3/\epsilon$ for all $t$.

Finally, because $O \leq M_t(x) \leq 1$ for all $t$ and $x$, 
the Hoeffding bound gives that taking the sample mean of
$M_t(x)$ over a sample of size $\Omega(\epsilon^{-2})$ will, with
constant probability, produce an estimate with additive error at most
$\epsilon/3$.  Furthermore, if the algorithm terminates in $T$ steps,
then a single uniform random sample $R$ of size $\Omega(\log(T)/\epsilon^2)$
guarantees, with constant probability, that estimating the expected
value of $M_t(x)$ by the sample mean over $R$ at every step $t$ will
produce an $\epsilon/3$ accurate estimate at every step.

Figure~\ref{figure:filtsmoothboost} presents the modified SmoothBoost
algorithm.  Notice that in place of a sample $S$ representing the
target function $f$, we are assuming that we are given a membership
oracle $MEM_f$.  We will subsequently consider quantum versions of
this algorithm and of the membership oracle.  For this reason, we show
the definitions of $M$ and $N$ as being over all of $\szon$, although
for a classical algorithm the only values that would actually be used
are those corresponding to $x \in R$.

\begin{figure}[t] 

\begin{tabbing}
{\bf Output:} \= www \= www \= \kill
{\bf Input:} \> Parameters  $0 < \epsilon < 1/2$, $0 \le \gamma < 1/2$ \\
             \> Membership oracle $MEM_f$ \\
             \> Weak learning algorithm {\tt WL} \\
\\
{\bf Output:} \> Hypothesis $h$
\end{tabbing}
\begin{algo}
\algitem{0} Draw uniform random sample $R$ of $\Omega(\log(\epsilon^{-1} \gamma^{-1})/\epsilon^2)$ instances $x$ and label using $MEM_f$
\algitem{0} $U_R$ $\equiv$ the uniform distribution over $R$
\algitem{0} $M_{1}(x) \equiv 1$, $\forall x \in \szon$
\algitem{0} $N_{0}(x) \equiv 0$, $\forall x \in \szon$
\algitem{0} $\theta \leftarrow \gamma / (2 + \gamma)$
\algitem{0} $t \leftarrow 1$
\algitem{0} {\bf while } $\Exp_{x \sim U_R}[M_{t}(x)] > 2\epsilon/3$ {\bf do} 
\algitem{1} $D_{t}(x) \equiv M_{t}(x)/(2^n \Exp_{x \sim U_R}[M_{t}(x)])$ 
\algitem{1} $h_{t} \leftarrow \mathtt{WL}(MEM_f,D_{t},
\delta = \Omega(\epsilon^{-1} \gamma^{-2}))$
\algitem{1} $N_{t}(x) \equiv N_{t-1}(x) + f(x)h_{t}(x) - \theta$, $\forall x \in \szon$
\algitem{1} $M_{t+1}(x) \equiv \iverson{N_{t}(x) < 0} + (1-\gamma)^{N_{t}(x)/2}\iverson{N_{t}(x) \ge 0}$,
        $\forall x \in \szon$
\algitem{1} $t  \leftarrow t + 1$
\algitem{0} {\bf end while}
\algitem{0} $T \leftarrow t - 1$
\algitem{0} $H \equiv \frac{1}{T}\sum_{i=1}^{T} h_{i}$
\algitem{0} {\bf return} $h \equiv \sign(H)$.
\end{algo}
\hrule

\caption[Modified SmoothBoost algorithm.]
{The {\tt SmoothBoost} modified for boost-by-filtering.}
\label{figure:filtsmoothboost}
\end{figure}

While the {\tt SmoothBoost} algorithm has been presented for
illustration, Klivans and Servedio \cite{ks99} have shown that one of
Freund's boosting algorithms, which they call 
{\tt B$_{\mbox{Comb}}$}, 
is actually slightly superior to {\tt SmoothBoost} for
our purposes.  Specifically, they note that {\tt B$_{\mbox {Comb}}$}
has properties similar to those of {\tt SmoothBoost} given in
Lemma~\ref{lem:smoothboost}, with the change that the number of stages
$T$ improves from $O(\epsilon^{-1} \gamma^{-2})$ to
$O(\log(1/\epsilon)/\gamma^2)$ while the smoothness of each of the
distributions $D_t$ passed to the weak learner satisfies (when
learning over all of $\szon$) $L_{\infty}(2^n D_t) =
O(\log(1/\epsilon)/\epsilon)$.  we will continue to use
{\tt SmoothBoost} in our analysis here, since 
{\tt B$_{\mbox {Comb}}$} and its analysis are noticeably more
complicated than {\tt SmoothBoost} and its analysis.  However,
our final sample size bounds will be stated as if {\tt B$_{\mbox {Comb}}$}
is being used, and the final version of this paper will include
details of the {\tt B$_{\mbox {Comb}}$} analysis.

\ignore{
It is worth noting that given the smoothness guarantee of
the boost-by-filtering version of {\tt SmoothBoost}, there is
a classical algorithm for weak learning DNF in time $\tilde{O}(ns^{4}/\epsilon^{2})$.
}

\section{A query-efficient quantum {\tt WDNF} algorithm}

In this section we describe a quantum weak learning algorithm {\tt WDNF} for finding parity
approximators of non-Boolean functions under smooth distributions. This algorithm is based on
a quantum Goldreich-Levin algorithm given by Adcock and Cleve \cite{ac01}.
For completeness we describe the quantum Goldreich-Levin algorithm in the following.
This algorithm utilizes the Pauli $X$ (complement) and $Z$ (controlled phase flip) gates and 
the Hadamard gate $H$ defined as follows.
\[
X = \begin{pmatrix} 0 & 1 \\ 1 & 0\end{pmatrix}, \ \ \
Z = \begin{pmatrix} 1 & 0 \\ 0 & -1\end{pmatrix}, \ \ \
H = \frac{1}{\sqrt{2}}\begin{pmatrix}1 & 1\\1 & -1\end{pmatrix}.
\]
Let $H_{n} = H^{\otimes n}$ ($n$-fold tensor of $H$ with itself) be the Walsh-Hadamard transform
on $n$ qubits and let $U_{0}(\sum_{x} \alpha_{x}|x\rip) = \sum_{x \neq 0}\alpha_{x}|x\rip - \alpha_{0}|0\rip$
be the unitary transformation that flips the phase of the all-zero state.

The $U_{MQ}$ transformation that represents a {\em noisy} membership oracle with respect to a parity
function $\chi_A$ defined in \cite{ac01} is given by
\[
        U_{MQ}|x\rip|0_m\rip = \alpha_x|x,u_x,A\cdot x\rip + \beta_x|x,v_x,\overline{A\cdot x}\rip,
\]
where $\sum_x \alpha_x^2 \ge 1/2+\gamma$ and $\sum_x \beta_x^2 \le 1/2-\gamma$. By a result of
Jackson \cite{j97}, for any DNF formula $f$ with $s$ terms, there is a parity function $A$ such that
$\Pr[f(x) = \chi_A(x)] \ge 1/2 + \gamma/2$, for $\gamma = 1/(2s+1)$. Thus, a noiseless DNF oracle
$QMQ_f$ is a noisy oracle $U_{MQ}$ for some parity function $\chi_A$.
Thus, we may assume that $U_{MQ}$ is a unitary  transformation that represents a quantum membership
oracle $QMQ_{f}$ for a DNF formula $f$ that maps $|x\rip|0_{m}\rip$ to $|x\rip|u_{x},f(x)\rip$, 
for some string $u_{x} \in \zo^{m-1}$ that represents the work space of the oracle. 

\begin{figure}[t] 

\begin{tabbing}
{\bf Output:} \= mm \= mm \= \kill
{\bf Input:} \> Parameters $n$,  $\gamma \in (0, 1/2)$, $\delta > 0$ \\
             \> Quantum membership oracle $QMQ_f$ for Boolean function $f$\\ 
             \> \> represented by a unitary tranformation $U_{MQ}$ \\
             \> Random uniform sample $R$ of size $\tilde{\Omega}(\gamma^{-2}\log(1/\delta))$.
\\
{\bf Output:} \> A coefficient $A$ with the property that $\Pr_D[f = \chi_{A}] \ge \frac{1}{2}+\gamma$ \\
            \> \> with probability at least $1-\delta$.
\end{tabbing}

\begin{algo}
\algitem{0} Let $C$ be defined as in Equation \ref{eqn:qgl}.
\algitem{0} Label $R$ using $QMQ_f$.
\algitem{0} Define a sampling-based $U_{EQ}$ as in Equation \ref{eqn:eq}.
\algitem{0} $|\phi\rip \leftarrow C|0_n\rip_{I}|0_m\rip_{A}|0\rip_{B}$
\algitem{0} {\bf for } $k = 1,\ldots,O(1/\gamma)$ {\bf do}
\algitem{1} $|\phi\rip \leftarrow -CU_{0}C^{\dagger}U_{EQ}|\phi\rip$
\algitem{0} {\bf end for}
\algitem{0} Measure and return the contents of register $I$.
\end{algo}
\hrule
\ignore{
\begin{algo}
\algitem{0} Prepare the initial superposition $|0_{n}\rip_{I}|0_{m}\rip_{A}|0\rip_{B}$ in the
        $I$ (input), $A$ (answer), and $B$ (bit) registers.
\algitem{0} Apply $H_{n}$ to register $I$ and $X$ to register $B$. 
\algitem{0} Apply $U_{MQ}$ to registers $I$ and $A$.
\algitem{0} Apply $Z$ to the last bit of $A$ (control) and $B$.
\algitem{0} Apply $U_{MQ}^{\dagger}$ to registers $I$ and $A$.
\algitem{0} Apply $H_{n}$ to register $I$.
\end{algo}
}

\caption[A quantum weak learning algorithm.]
{The quantum weak learning algorithm {\tt QWDNF} for uniform distribution.}
\label{figure:qac}
\end{figure}

The quantum algorithm {\tt QGL} of Adcock and Cleve is 
represented by the following unitary transformation
\begin{equation} \label{eqn:qgl}
        C = (H_{n} \otimes I_{m+1}) (U_{MQ}^{\dagger} \otimes I_{1}) 
                (I_{n+m-1} \otimes Z) (U_{MQ} \otimes I_{1}) (H_{n} \otimes I_{m} \otimes X)
\end{equation}
applied to the initial superposition of $|0_n,0_m,0\rip$. 
In \cite{ac01} it was proved that the quantum algorithm {\tt QGL} prepares a superposition of 
all $n$-bit strings such that the probability of observing the coefficient $A$ is $4\gamma^{2}$. 
By repeating this for $O(1/\gamma^{2})$ stages, we can recover $A$ with constant probability.
 
The number of stages can be reduced to $O(1/\gamma)$ by using a technique called 
{\em amplitude amplification}. This amplification technique uses an iterate of the form
\[
        G = (-C U_{0}C^{\dagger}U_{EQ})^{k} C|0_{n},0_{m},0\rip,
\]
where 
$k$ is approximately $O(1/\gamma)$, and $U_{EQ}$ is a unitary transformation that represents
a {\em quantum equivalence} oracle $QEQ_{f}$. The transformation $U_{EQ}$ is defined as
\begin{equation} \label{eqn:eq}
        U_{EQ}|a\rip = \left\{\begin{array}{ll}
                                -|a\rip & \mbox{ if $|\Exp[f\chi_{a}]| \ge \theta$} \\
                                |a\rip  & \mbox{ otherwise }
                                \end{array}\right.
\end{equation}
For the purpose of learning DNF, we need to simulate $U_{EQ}$ using a
sampling algorithm that has access to $QMQ_{f}$. A classical
application of Hoeffding sampling requires $\Omega(1/\gamma^{2})$
queries to $QMQ_{f}$\footnote{Grover has proposed a quantum algorithm
  for estimating the mean that requires
  $O(\frac{1}{\gamma}\log\log\frac{1}{\gamma})$ queries.  However, in
our setting, we will use fewer queries if we estimate this value
classically because we can use a single sample for all estimates, as
discussed below.}. To simulate $U_{EQ}$, we will simply use a sample $R$ of size
$\Omega(1/\gamma^2\log(1/\delta))$ to obtain a good estimate with probability at
least $1-\delta$.

Finally, recall that we will be applying boosting to this weak
learning algorithm, which means that {\tt QWDNF} will be called a
number of times.  However, it is not necessary to draw a new random
sample $R$ each time {\tt QWDNF} is called, as the boosting algorithm
merely wants a guarantee that the algorithm succeeds with high probability
and does not require independence.
The resulting quantum weak learning algorithm
for DNF, which we denote {\tt QWDNF},
is described in Figure~\ref{figure:qac}.

\subsection{Non-Boolean Functions over Smooth Distributions}

Recall that in the Harmonic Sieve algorithm \cite{j97}, we need to find weak Parity approximators
for non-Boolean functions $g$ that is based on the DNF formula $f$ and the current boosting distribution
$D$ in {\tt SmoothBoost}, i.e., we need to consider expressions of the form 
(we have dropped subscripts for convenience)
\begin{eqnarray*}
\Exp_{D}[f(x)\chi_{A}(x)] & = & \sum_{x} D(x)f(x)\chi_{A}(x) \\
                & = & \sum_{x} \frac{M(x)}{2^{n}\Exp[M(x)]}f(x)\chi_{A}(x) \\
                & = & \frac{\Exp[M(x)f(x)\chi_{A}(x)]}{\Exp[M(x)]}.
\end{eqnarray*}
This shows a reduction from finding a coefficient $A$ such that $|\Exp_D[f\chi_A]|$ is large to
finding a coefficient $A$ so that $|\Exp_U[g\chi_A]|$, where $g(x) = M(x)f(x)$, is large.
Assuming that $\Exp[M(x)] \ge \epsilon/3$, we will use the 
algorithm {\tt QWDNF} to find a coefficient $A$ 
such that for some constant $c_2$ 
\[
                |\Exp[M(x)f(x)\chi_{A}(x)]| \ge \frac{c_2 \epsilon}{3(2s+1)} \dotdef \Gamma.
\]

Note that $0 < M(x) \le 1$, for all $x$. Thus we can use a technique of Bshouty and Jackson 
\cite{bj99} that transforms the problem to the individual bits of $M(x)$. 
Let $d = \log(3/\Gamma)$, where $\Gamma$ is as above. 
Let $\alpha(x) = \lfloor 2^{d}M(x)\rfloor/2^{d}$, i.e., $M(x)$ truncated to include only $d$ of
its most significant bits. Assume that $\alpha = \sum_{j=1}^{d} \alpha_{j}2^{-j} + k2^{-d}$, where
$\alpha_{j} \in \{-1,1\}$ and $k \in \{-1,0,1\}$. Thus
\[
        |\Exp[M(x)f(x)\chi_{A}(x)]| - \frac{\Gamma}{3} 
        \le
        |\Exp[\alpha(x)f(x)\chi_{A}(x)]| 
        \le
        \max_{j} |\Exp[\alpha_{j}(x)f(x)\chi_{A}(x)]| + \frac{\Gamma}{3}
\]
thus there exists $j$ so that $|\Exp[\alpha_{j}(x)f(x)\chi_{A}(x)]| \ge \Gamma/3$, assuming
$|\Exp[M(x)f(x)\chi_{A}(x)]| \ge \Gamma$.

Note that to simulate $U_{EQ}$ for verifying that the non-Boolean function $g(x) = M(x)f(x)$ has
a $\Gamma$-heavy coefficient at $A$, i.e., $|\hat{g}(A)| \ge \Gamma$, we need a sample of size
at least $1/\Gamma^2 \sim (s/\epsilon)^2$.

\section{A quantum Harmonic Sieve algorithm}

In this section, we describe a quantum version of the Harmonic Sieve
algorithm obtained by combining the quantum Goldreich-Levin algorithm
and the {\tt SmoothBoost} boosting algorithm (see Figure~\ref{figure:hs}).

The top level part of this algorithm involves $O(s^2/\epsilon)$
boosting rounds\footnote{This could be improved to
  $O(s^2\log(1/\epsilon))$ rounds if Freund's {\tt BComb} algorithm is
  used.} and each round requires invoking the algorithm {\tt QWDNF}
that uses $\tilde{O}(s/\epsilon)$ queries.  The ``oracle''
$QMQ_f \cdot D_t$ represents the procedure that will produce Boolean
functions representing the bits of $M_t f$ and simulate quantum
membership oracles to be passed to {\tt QWDNF}.
  There is an additional cost of a
random sample of size $\tilde{O}(s^2/\epsilon^2)$ for estimating the
expression $\Exp[M_t]$ to within $O(\epsilon)$ and for simulating the
equivalence oracle $U_{EQ}$ used by {\tt QWDNF}. The latter step
requires estimating the expression $\Exp[M_tf\chi_{A}]$ to within
$O(\epsilon/s)$ accuracy. This random sample is shared among all
boosting stages and all calls to {\tt QWDNF}. The key property
exploited here is the oblivious nature of the sampling steps.

Thus the overall algorithm, if {\tt BComb} is used as the boosting
algorithm, requires $\tilde{O}(s^3/\epsilon + s^2/\epsilon^2)$ sample 
complexity.  The best classical algorithm (also based on {\tt BComb})
has complexity $\tilde{O}(ns^2/\epsilon^2)$.  Thus, for 
$s = \Theta(1/\epsilon)$, the quantum algorithm is an improvement
by a factor of $n$.

\ignore{\footnote{The sample complexity is $\tilde{O}(s^3 + s^2/\epsilon^2)$ with {\tt BComb}.}.}

\begin{figure}[t] 

{\bf Input:} Parameters  $0 < \epsilon,\delta < 1$, 
        $n$, 
        a quantum membership oracle $QMQ_{f}$ for a DNF formula $f$,
        $s$ (the size of DNF $f$),

{\bf Output:} $h$ so that $\Pr[f \neq h] < \epsilon$.

\begin{algo}
\algitem{0} Draw a uniform random sample $R$ of $\Omega(s^2/\epsilon^2)$ instances $x$ and label using $QMQ_f$
\algitem{0} $\gamma \leftarrow 1/(8s+4)$ (weak advantage)
\algitem{0} $k \leftarrow c_{1}\gamma^{-2}\epsilon^{-1}$ (number of boosting stages)
\algitem{0} $M_{1} \equiv 1$ (all-one function)
\algitem{0} $N_{0} \equiv 0$ (all-zero function)
\algitem{0} {\bf for } $t = 1,\ldots,k$ {\bf do}
\algitem{1} $E_{t} \leftarrow \Exp_{x \sim U_R}[M_{t}(x)]$ 
\algitem{1} {\bf if } $E_{t} \le 2\epsilon/3$ {\bf then} 
\algitem{2} {\bf break} 
\algitem{1} {\bf end if}
\algitem{1} $D_{t} \equiv M_{t}/(2^{n}E_{t})$
\algitem{1} $h_{t} \leftarrow \mathtt{QWDNF}(n,\gamma \epsilon,\delta/2k,QMQ_{f}\cdot D_{t},R)$ where
$\Pr_{D_{t}}[h_{t}(x) \neq f(x)] \le \frac{1}{2}-\gamma$.
\algitem{1} $N_{t} \equiv N_{t-1} + fh_{t} - \theta$
\algitem{1} $M_{t+1} \equiv \iverson{N_{t} < 0} + (1-\gamma)^{N_{t}/2}\iverson{N_{t} \ge 0}$
\algitem{0} {\bf end do}
\algitem{0} $T = t - 1$
\algitem{0} $H(x) \equiv \frac{1}{T}\sum_{i=1}^{T} h_{i}(x)$
\algitem{0} {\bf return} $h(x) = \sign(H(x))$.
\end{algo}
\hrule

\caption[Quantum Harmonic Sieve algorithm.]
{The new {\tt QHS} algorithm.}
\label{figure:hs}
\end{figure}

\section{Lower bounds}

In this section, we prove a lower bound on the query complexity of any quantum PAC learning algorithm 
for DNF formulae.

\begin{theorem}
Let $s \ge n/\log n$.
Then any quantum PAC learning algorithm requires $\Omega(s\log n/n)$ queries to learn a DNF 
formula of size $s$ over $n$ variables under the uniform distribution, 
given $\epsilon < 1/4$ and any constant $\delta > 0$.
\end{theorem}
\prf
We use a construction given in Bshouty et al. \cite{bjt99}.
Let $t = \log s$ and $u = n - t$. Consider the following class $C$ of DNF formulae over the variable 
set of $V = \{x_1,\ldots,x_t\} \cup \{y_1,\ldots,y_u\}$,
\[
        C = \left\{\bigvee_{a \in \zo^{t}} x^{a}y_{a} \ : \ \lip y_{a}\rip_{a \in \zo^{t}}\right\},
\]
where $x^{a} = \bigwedge_{i=1}^{t} x_{i}^{a_{i}}$, with the convention $x_i^0 = x_i$ and $x_i^1 = \overline{x_i}$,
and for each $a \in \zo^t$, $y_{a}$ is a constant (0 or 1) or one of the variables $y_i$ or its negation.
Each $f \in C$ is specified uniquely by a word $y \in \Sigma^{s}$ over the alphabet
$\Sigma = \{0,1,y_1,\overline{y_1},\ldots,y_u,\overline{y_u}\}$, i.e., we may denote $f_y$ to be the DNF
specified by the word $y \in \Sigma^s$. By the Gilbert-Varshamov bound, there is a code $L \subset \Sigma^{s}$ 
with minimum distance $\alpha s$ of size at least 
\[
        \frac{|\Sigma|^{s}}{\sum_{k=0}^{\alpha s} \binom{s}{k}(|\Sigma|-1)^{k}}
        \ge
        \left(\frac{(2u+2)^{1-\alpha}}{2}\right)^{s}
\]
We focus on $C_L \subset C$ where the words $y$ are taken from $L$.
Note that for any distinct $y,z \in L$ we have $\Pr_U[f_y \neq f_z] = \Exp_U[f_{y \oplus z}] \ge \alpha s/2$, 
where the probability is taken over the uniform distribution on $V$. Letting $2\epsilon = \alpha s/2$, this
implies that any two distinct DNF functions $f_y,f_z$, where $y,z \in L$, are $(2\epsilon)$-separated. 
So any $(\epsilon,\delta)$-PAC algorithm for $C_L$ must return exactly the unknown target function.

Now let $A$ be any quantum $(\epsilon,\delta)$-PAC algorithm with access to a quantum membership oracle 
$QMQ_{f}$ associated with a target DNF function $f$. Suppose that $A$ makes $T$ queries for any function
$f \in C_L$. Following the notation in \cite{gs01}, let $X^f$ be the truth table of the DNF function $f$, 
i.e., $X^f$ is a binary vector of length $N = 2^n$. Let $P_{h}(X^{f})$ be the probability function of $A$ 
of returning as answer a DNF function $h$ when the oracle is $QMQ_f$, for $h,f \in C_L$. 
By the PAC property of $A$, we have
\begin{itemize}
\item $P_f(X^f) \ge 1 - \delta$
\item $\sum_{h: h \neq f} P_h(X^f) < \delta$
\end{itemize}
It is known that $P_f$ is a multivariate polynomial of degree $2T$ over $X^h$, for any $f,h$. 
Let $N_0 = \sum_{t=0}^{2T} \binom{N}{t}$. For $X \in \zo^N$, let $\tilde{X} \in \zo^{N_0}$ be the vector 
obtained by taking all $\ell$-subsets of $[N]$, $\ell \le 2T$. The coefficients of $P_h$ can be specified
by a real vector $V_h \in \Real^{N_0}$ and $P_h(X^f) = V_h^{T}X^f$. 
Let $M$ be a matrix of size $|C_L| \times N_0$ whose rows are given by the vectors $V_h^{T}$ for all $h \in C_L$. 
Let $N$ be a matrix of size $|C_L| \times |C_L|$ whose columns are given by the vectors $MV_g$ for all $g \in C_L$.
Observe that the $(h,f)$ entry in the matrix $N$ is given by $P_h(X^f)$.
As in \cite{gs01}, we argue that since $N$ is {\em diagonally dominant} (from the PAC conditions on $\delta$ above), 
it has full rank. Thus $N_0 \ge |C_L|$, which implies that 
\[
        N^{2T} \ge |C_L| \ge \left(\frac{(2u+2)^{1-\alpha}}{2}\right)^{s}.
\]
This implies that $4nT \ge s\log(n)(1-o(1))$ which gives $T \ge \Omega(s\log n/n)$.
\qed

\ignore{\section{Open Questions}

An immediate question is whether there are stronger lower bounds for quantum PAC learning of DNF.
Another question is if there is a pure quantum mechanical algorithm for boosting in the PAC model.
}

\section{Acknowledgments}

The second author thanks Richard Cleve for helpful discussions on the quantum Goldreich-Levin algorithm.

\end{document}